\documentclass[twocolumns]{aastex63}
\usepackage{lineno}
\usepackage[caption=false]{subfig}
\usepackage{multirow}
\usepackage{graphicx}
\usepackage{booktabs}
\usepackage{amsmath,amssymb}
\usepackage[T1]{fontenc}

\graphicspath{{./}{figures/}}

\begin{document}

\title{Quasi-periodic Oscillations in GRB 210514A: a Case of a Newborn Supra-Massive Precessing Magnetar Collapsing into a Black Hole?}
\correspondingauthor{Le Zou}
\email{zoule@xtu.edu.cn}
\author{Le Zou}
\affil{Department of Physics, Xiangtan University, Xiangtan 411105, People's Republic of China}
\affil{Key Laboratory of Stars and Interstellar Medium, Xiangtan University, Xiangtan 411105, People's Republic of China}
\author{Ji-Gui, Cheng}
\affil{School of Physics and Electronics, Hunan University of Science and Technology, Xiangtan 411201, People's Republic of China}

\begin{abstract}
Magnetar is proposed as one of the possible central engines for a gamma-ray burst (GRB). Recent studies show that if a magnetar has a rotational axis misaligned from the magnetic one, a periodic lightcurve pattern is expected with a period of seconds to minutes. Inspired by this unique feature, in this paper, we search for the quasi-periodic oscillation (QPO) signals in the {\it Swift} observations of GRBs. Using the Lomb-Scargle periodogram and the weighted wavelet Z-transform algorithms, we find that the {\it Swift}-BAT data of GRB 210514A has a QPO signal with a period $\sim 11\,{\rm s}$. The estimated confidence level of the signal is over $3 \sigma$. The global lightcurve of this GRB exhibits a double-plateau structure with a sharp decay segment between plateaus. The lightcurve feature resembles those of GRBs that were reported to have internal plateaus. We explain the observations of GRB 210514A with a supra-massive magnetar (SMM) model, where the QPO signal in the first plateau is produced via the dipole radiation of the SMM experiencing a precession motion, the sharp decay is due to the collapse of the SMM into a black hole (BH), and the second plateau could be produced via the fall-back accretion of the newborn BH. We fit the precession model to the observations using the Bayesian statistic and the best-fit magnetar parameters are discussed. Alternative models concerning a BH central engine may also provide reasonable explanations for this burst, only in this case the QPO signal could merely be a coincidence.
\end{abstract}


\section{Introduction}
\label{sec:intro}
{
Gamma-ray bursts (GRBs) are among the most energetic events observed in our universe. It is widely believed that a GRB is associated either with the death of a massive star or the coalescence of two compact objects, i.e. a compact remnant will be born in these two processes and serves as the central engine powering the GRB (e.g. \citealt{2007ApJ...655..989Z, 2015PhR...561....1K}). Some researches suggest that this remnant could be a black hole (BH) with an efficient accretion disk \citep{1992ApJ...395L..83N, 1999ApJ...524..262M, 2001ApJ...557..949N}, or a rapidly rotating magnetar \citep{1992Natur.357..472U, 1994MNRAS.270..480T, 1998PhRvL..81.4301D, 1998A&A...333L..87D, 2000ApJ...537..810W,  2011MNRAS.413.2031M, 2014ApJ...785...74L}. To know exactly what remnant is produced has long been an open question in GRB studies, and this issue is intensively discussed for the famous GW170817/GRB 170817A event, which is attributed to a merger of two neutron stars \citep{2017ApJ...848L..12A, 2017PhRvL.119p1101A}. One plausible way to discriminate between the BH and the magnetar central engines of GRBs is to estimate the total energy. Since the energy release of a magnetar is strictly constrained by its initial rotational energy, a BH central engine would be favored if a GRB has a total energy exceeding the maximum energy budget of a magnetar (e.g. \citealt{2014ApJ...785...74L}). Nevertheless, estimating the total energy could be a real challenge in most observed GRBs and the jet opening angle is difficult to acquire in their afterglow data.

An alternative method is to find particular lightcurve features within the observations of GRBs. For instance, the X-ray shallow decay or external plateau observed in lightcurves is attributed to the spindown of a magnetar \citep{1998PhRvL..81.4301D, 1998A&A...333L..87D, 2001ApJ...552L..35Z}. Such features have already been observed in the {\it Swift}-XRT data of some GRBs \citep{2014ApJ...785...74L, 2018MNRAS.480.4402L}. However, these signatures do not necessarily imply the presence of a magnetar, as matters ejected by a BH central engine can create a continuous energy injection even if they are ejected within a relatively short time frame \citep{1998ApJ...496L...1R, 2007ApJ...665L..93U, 2007MNRAS.381..732G, 2012ApJ...761..147U}. A smoking gun signature of a magnetar central engine emerged with the discovery of the internal plateau \citep{2007ApJ...665..599T, 2007ApJ...670..565L, 2010MNRAS.409..531R, 2014ApJ...785...74L, 2015ApJ...805...89L, 2019ApJ...877..153Z, 2021MNRAS.508.2505Z}. This plateau is distinct from external shock model prediction mainly due to the existence of a steep decay at the end of the plateau, which reflects a sudden cessation of the central engine. Within the magnetar model, this is interpreted as the supra-massive magnetar (SMM), i.e. the central engine, collapses into a BH after significantly dissipating its rotational energy \citep{2014ApJ...780L..21Z}. However, the internal plateau feature is extremely rare in GRB observations as the spindown luminosity is expected to be relatively smooth lacking significant fluctuations \citep{1992Natur.357..472U,1994MNRAS.267.1035U} and the observed prompt emission of GRBs are often erratic with many pulses.

Interestingly, recent researches suggest that the quasi-periodic oscillations (QPO) signal superimposed on the spindown luminosity of GRBs might originate from the precession motion of a magnetar, serving as additional evidence for a magnetar central engine \citep{2020ApJ...892L..34S, 2021ApJ...921L...1Z, 2022MNRAS.513L..89Z}. This specific precession motion is caused by the misalignment between the magnetar's rotational axis and its magnetic axis \citep{2005ApJ...634L.165S, 2007Ap&SS.308..119D, 2014MNRAS.441.1879P, 2019ApJ...886....5S}. Assuming a high ellipticity $\epsilon \sim 10^{-4}$ and an initial spin period $P_{\rm s,0} \sim 1\;{\rm ms}$, this motion could modulate GRB lightcurves with an evolving QPO signal. The precession period $P_{\rm p} = P_{\rm s} / \epsilon$ would thus range from tens to several hundred seconds. Our previous searches have identified QPO signals in two specific GRBs. In GRB 101225A, two plausible QPO signals with periods of $488\;{\rm s}$ and $\sim 250-300\;{\rm s}$ were detected \citep{2021ApJ...921L...1Z}. In GRB 180620A, a QPO signal with a period of $\sim 650\;{\rm s}$ was observed \citep{2022MNRAS.513L..89Z}. These signals in both bursts were found in the X-ray plateaus of their afterglow lightcurves, which are consistent with the magnetar precession model. If the magnetar central engine truly exists, it is reasonable to expect that more cases have already been observed \citep{2020ApJ...894...52X}. However, identifying a QPO signal is not an easy task that requires a delicate data analysis \citep{2023Natur.613..253C}.

By systematical analysis of the accumulated {\it Swift} data from 2005 to 2024, we find a peculiar GRB 210514A, which potentially has a QPO signal in the {\it Swift}-BAT observation. This burst is one of few observed GRBs having a double-plateau lightcurve structure (see e.g. \citealt{2018MNRAS.475..266Z, 2020ApJ...896...42Z}). Its first plateau segment appears in the BAT observations and consists of continuous multi-flares ranging from $0$ to $\sim 85$~s. The second plateau segment is shown in the XRT data ranging from $\sim 100$ to $1000$~s. Between the two plateau segments, there is a significant flux decrease. Such lightcurve segments can be reconciled with an SMM central engine, where the first plateau can be explained as the internal plateau, the rapid flux decrease is due to the collapse of the SMM into a BH, and the second plateau can be produced via the fall-back accretion of the newborn BH \citep{2007ApJ...665..599T, 2018MNRAS.475..266Z, 2020ApJ...896...42Z}. In this scenario, we propose that the continuous multi-flares in the first plateau segment are produced by the precession motion of an SMM. In this work, we perform a detailed time-series analysis on the lightcurve of GRB 210514A aiming to explore its physical origin. The paper is organized as follows. In Sec.~\ref{sec:data}, we collect the {\it Swift} observations of GRB 210514A and conduct the data analysis. In Sec.~\ref{sec:mdl}, we describe our model and its application. Discussions and conclusions are provided in Sec.~\ref{sec:dis} and Sec.~\ref{sec:con}. Throughout this paper, a flat-$\Lambda$CDM cosmological model is used with parameters $H_{0} = 70$ ${\rm km}$ ${\rm s}^{-1}$ ${\rm Mpc}^{-1}$, $\Omega_{M}=0.3$, and $\Omega_{\Lambda}=0.7$.

\section{DATA ANALYSIS}
\label{sec:data}
{
Based on the historical observations of the {\it Swift} satellite \citep{2010A&A...519A.102E}\footnote{The {\it Swift} observational data are available at the UKSSDC website: \url{https://www.swift.ac.uk}.}, we search for potential QPO signals in GRBs. In this process, we adopt a Bayesian method, i.e. the generalized Lomb-Scargle periodogram (LSP) algorithm \citep{1976Ap&SS..39..447L, 1982ApJ...263..835S}, to derive the power spectral density (PSD) in the frequency domain of GRB lightcurves. The time-series analysis is implemented with the Python {\tt astropy} module. Following the analysis procedure introduced in \cite{2010MNRAS.402..307V} and \cite{2017A&A...600A.132S}, we fit the noise in the derived PSD with a power-law function. Then the single-frequency and multi-frequency (global) false-alarm probabilities are estimated. Our results show that only a few GRBs (including the previously reported GRB 101225A and 180620A) present a tentative QPO signal, in which only GRB 210514A's prompt emission shows a QPO signal and the highest PSD peak at $\sim 11$~s exceeding the global 99.7\% confidence level as presented in the Fig.~\ref{fig:QPO}(a). It should be noted that the BAT data are extrapolated to the XRT energy band of $[0.3-10]\;{\rm keV}$ \citep{2006ApJ...647.1213O, 2007A&A...469..379E, 2009MNRAS.397.1177E}. The QPO signal in GRB 210514A could be a possible candidate. To further confirm that the candidate is real, we use the weighted wavelet Z-transform (WWZ) algorithm \citep{1996AJ....112.1709F} to calculate the PSD map in the time-frequency domain. The result is shown in Fig.~\ref{fig:QPO}(b), which also suggests that a PSD peak is located around $\sim 11$~s with no significant evolution with time. Hence, we propose that the $P \sim 11$~s QPO signal is identified in GRB 210514A.

GRB 210514A is a long GRB with $T_{\rm 90} \approx 70.2\;{\rm s}$ and its total gamma-ray outburst lasts $\sim 80$~s \citep{2021GCN.30009....1L, 2021GCN.30018....1M}. It was first triggered by the BAT (the trigger time is $642709453.824$~s in MET) and then observed by the XRT starting from $t_{\rm obs} \sim 60\;{\rm s}$. By extrapolating the BAT data to the XRT energy band, the global lightcurve of this burst has a good consistency that the BAT data is overlapping with the XRT ones in $t_{\rm obs} \sim [40-50]\;{\rm s}$, as shown in the Fig.~\ref{fig:lightcurve}, suggesting the same origin of these data. Meanwhile, there are four distinct power-law segments which consist of the global lightcurve. We fit them with two broken power-law (BPL) functions. The BPL function is written as
\begin{equation}
F_{\rm obs} = F_{\rm obs,0} \left[ \left( \frac{t_{\rm obs}}{t_{\rm obs,b}} \right)^{p_{\rm 1} w} + \left( \frac{t_{\rm obs}}{t_{\rm obs,b}} \right)^{p_{\rm 2} w} \right]^{-1/w},
\label{eq:bpl}
\end{equation}
where $w$ describes the sharpness of the break, $p_{\rm 1}$ and $p_{\rm 2}$ are the decay indices before and after $t_{\rm obs,b}$, respectively. During fitting, we estimate the reduced-$\chi^{2}$ value to indicate the goodness of the fit, and the results are summarized in Tab.~\ref{tab:fitting}. Since the index $p_{\rm 1}$ for both of the time intervals are lower than 0.75 \citep{2007ApJ...670..565L, 2008ApJ...675..528L}, we will refer to the corresponding lightcurve segments as the first and second plateaus, respectively. Note that the fit goodness in the first plateau is not as good as the second one. This is probably caused by the complex flux variability during $t_{\rm obs} \sim [0, 80]\;{\rm s}$, which will be discussed later in the Sec.~\ref{sec:mdl}.

As for the host galaxy of GRB 210514A, unfortunately, there are no relevant observations have been reported and the redshift of this burst is currently unknown. So we have to use a pseudo-redshift instead to calculate the luminosity. Such a redshift can be estimated via empirical correlations established upon redshift-known GRBs. \cite{2023ApJ...943..126D} derive the pseudo-redshift of GRB 210514A to be 0.57 through the relation between the X-ray luminosity $L_{\rm X}$, the end time of the plateau segment $T_{\rm a}$, and the isotropic gamma-ray energy $E_{\rm \gamma, iso}$. Therefore, we will use that value throughout this paper.
}

\section{Model and Application}
\label{sec:mdl}
{
Here, we propose that the peculiar global lightcurve of GRB 210514A is related to a newborn SMM formed with the burst. To be supra-massive, the mass of the magnetar should be in the range of $\left[ M_{\rm TOV}, 1.2M_{\rm TOV} \right]$ \citep{2017ApJ...850L..19M}, where $M_{\rm TOV}$ is the maximum mass allowed for non-spinning neutron stars and its value depends on which equation-of-state (EoS) model is adopted. In this work, we adopt a widely used EoS model GM1 with $M_{\rm TOV}=2.37M_{\odot}$ \citep{2014PhRvD..89d7302L, 2015ApJ...805...89L, 2016PhRvD..93d4065G}. Owing to the excessive weight, the initial spin period of the magnetar must be fast, so that its large angular momentum can prevent itself from collapsing into BH immediately after the birth. However, the magnetar eventually collapses as its rotational energy dissipates. Some of the energy is converted into electromagnetic (EM) emissions via the dipole radiation (DR) mechanism and causes the first plateau segment. Others may convert to gravitational wave (GW) emission whose intensity depends on the ellipticity of the magnetar. When the spin is slowing down, the magnetar experiences strong precession motion since its rotational axis deviates from its magnetic axis, producing a QPO signal in the first plateau segment. This dissipation process is short and several tens seconds later, the magnetar loses significant rotational energy that can not support its mass and collapses into a BH. The collapse of the magnetar suddenly interrupts the energy supply for the GRB and thus causes the steep decay segment. The newborn BH may gather the remnant material of the magnetar and provide a long-lasting energy injection into the GRB jet via the Blandford-Znajek (BZ) process \citep{1977MNRAS.179..433B}. This fall-back accretion results in the second plateau segment. As the surrounding materials are consumed, the accretion stops at $t_{\rm obs} \sim 1000$~s, and finally the GRB afterglow begins to be prominent, which explains the normal decay segment observed in the global lightcurve.

In this physical scenario, we are especially interested in the early evolution, i.e. the precession motion of the SMM. We introduce the relevant model details and formalism in the next subsection.
}

\subsection{The Precession Motion of a Newborn Supra-Massive Magnetar}
\label{subsec:mdl_precession}
{
In the general theory of magnetar spindown \citep{2001ApJ...552L..35Z, 2014PhRvD..89d7302L}, the luminosity evolution of the isotropic DR can be expressed as
\begin{equation}
L_{\rm X}(t)= L_{\rm k,0}\frac{\eta_{\rm X}}{f_{\rm b}}(1+\frac{t}{\tau_{\rm sd}})^{-\beta},
\label{eq:L}
\end{equation}
where $t = t_{\rm obs,b} / (1 + z) + \Delta t$ and $\Delta t$ is the delay time between the birth time of the SMM and the BAT trigger,$L_{\rm k,0}$ is the initial kinetic luminosity of the magnetar, $\eta_{\rm X}$ is the radiation efficiency, $f_{\rm b}$ is a correction factor for the beaming effect, and $\tau_{\rm sd}$ marks the characteristic spindown time scale. Note that we assume $\eta_{\rm X}/f_{\rm b} = 1$ for generality \citep{2013MNRAS.430.1061R, 2015ApJ...805...89L}, since its true value are related to the detailed jet production and dissipation process, which we don't have a comprehensive understanding yet. As suggested by the equation, the luminosity decay is slow for $t \lesssim \tau_{\rm sd}$ that behaves like a plateau, and the decay becomes obvious with slope $\beta$ after $t = \tau_{\rm sd}$. Typically, the value of $\beta$ is $2$ or $1$ depending on which radiation component dominates the spindown process. The value $\beta = 2$ corresponds to the DR dominated cases and $\beta = 1$ is for the GW emission dominated ones. In the case of GRB 210514A, i.e. SMM collapses before its spindown time scale \citep{2007ApJ...665..599T, 2018MNRAS.475..266Z}, $\beta$ cannot be determined by observations. Thus, we assume that the spindown is DR dominated,  i.e.  $\beta = 2$. The initial kinetic luminosity and spindown time scale can be estimated with magnetar parameters denoted as $\theta_{\rm m} = \left\{ B_{\rm p}, P_{\rm s, 0}, R, I \right\}$, in which $B_{\rm p}$ is the dipole magnetic field strength, $P_{\rm s, 0}$ is the initial spin period, $R$ is the radius, and $I$ is the inertial moment. The equations are
\begin{equation}
L_{\rm k,0,49} = 1.0 \times {B^2_{\rm p,15}P^{-4}_{\rm s,0,-3}R^{6}_{6}},
\label{eq:L_k_0}
\end{equation}
and
\begin{equation}
\tau_{\rm sd,3}=2.05 \times I_{45}B^{-2}_{\rm p,15}P^{2}_{\rm s,0,-3}R^{-6}_{6}.
\label{eq:tau_sd}
\end{equation}
In the Eqs.~\ref{eq:L_k_0} and \ref{eq:tau_sd}, quantities are written as $Q_{\rm m} = Q / 10^{\rm m}$ in the cgs units, e.g. $L_{\rm k, 0, 49} = L_{\rm k, 0} / (10^{49}\;{\rm erg/s})$.

The magnetar precession motion can induce a flux oscillation in the DR luminosity with a dimensionless factor $\lambda$, i.e. $L_{\rm p}(t) = L(t) \lambda$. Theoretically, $\lambda$ reflects the magnetosphere of the magnetar and its value depends on the magnetar orientation \citep{1969ApJ...157..869G, 2006ApJ...648L..51S, 2009A&A...496..495K, 2014MNRAS.441.1879P}. In this work, we adopt a hybrid model of $\lambda \approx 1 + \delta \sin^{2}(\alpha)$, where $\alpha$ is the inclination angle between the rotational and the magnetic axes, and $\delta$ quantifies the magnetospheric physics ($|\delta|\leq 1$) \citep{2014MNRAS.441.1879P, 2015MNRAS.453.3540A, 2020ApJ...892L..34S}. In the case of an SMM, the magnetic ``hair'' has to be ejected when it collapses into a BH according to the ``no hair'' theorem \citep{2014A&A...562A.137F, 2014ApJ...780L..21Z}. Thus, $\delta$ should be evolved with time. We parameterize this evolution as $\delta \approx -(b + t/\tau_{\rm c})^{a}$ \citep{2021ApJ...921L...1Z}, in which $\tau_{\rm c}$ is the collapse time scale of the SMM. Meanwhile, the evolution of $\alpha$ can be expressed as $\dot{\alpha} \approx k \Omega_{\rm p} \csc \alpha \sin(\Omega_{\rm p} t)$, where $\Omega_{\rm p}$ is the angular velocity of the precession motion and $k$ is a constant and is related to the other Euler angles \citep{1970ApJ...160L..11G, 2015MNRAS.451..695Z, 2020ApJ...892L..34S}. Given the initial inclination angle $\alpha_0$, we have
\begin{equation}
\begin{split}
\lambda = \lambda(t, \theta_{\rm p}) \approx & \left \{1 + \delta -\delta \left \{\left [ \alpha_{0} + k( \cos(\Omega_{\rm p}\times t)) \right ]^{2} \right \}\right \}\\
& \times \Big\{ 1 + \frac { t \left [ 1 + \delta \left( 1 - \alpha_{0}^2 - \frac {1} {2} k^2 \right) \right ] } {\tau_{\rm sd}} \\
&- \frac {k \delta \left[ 2 \alpha_{0} + \frac{1}{2} k \cos \left( \Omega_{\rm p} \times t \right) \right] \sin \left( \Omega_{\rm p} \times t \right)} {\tau_{\rm sd} \Omega_{\rm p}} \Big\}^{-2},
\label{eq:lambda}
\end{split}
\end{equation}
where $\theta_{\rm p} = \left\{ a, b,\alpha_0,k,\Omega_{\rm p} \right\}$ represents the parameters of the precession motion. Finally, the DR luminosity ($L_{\rm mp, X}$) considering the precession motion can be given by
\begin{equation}
L_{\rm mp, X}(t, \eta_{\rm X}, f_{\rm b}, \theta_{\rm m}, \theta_{\rm p}) = L_{\rm X}(t, \eta_{\rm X}, f_{\rm b}, \theta_{\rm m}) \lambda(t, \theta_{\rm p}),
\label{eq:L_mp_X}
\end{equation}
and in total $11$ parameters are required in this model. By fitting the model with the actual observations, one can derive the parameters and investigate the physical properties of the supra-massive magnetar.
}
}

\subsection{Applications to GRB 210514A}
\label{subsec:mdl_appli}
{
Using the magnetar precession model described in Sec.~\ref{subsec:mdl_precession}, we calculate the model line and fit it to the first plateau segment data in the global lightcurve of GRB 210514A. During fitting, the observed flux is converted to the isotropic luminosity in the comoving frame of cosmic expansion through $L_{\rm X,obs}=4\pi d_{\rm L}^{2} F_{\rm X, obs} \times (1+z)^{\Gamma-2}$ \citep{2001AJ....121.2879B, 2019ApJ...886....5S}, where $d_{\rm L}$ is the luminosity distance and $\Gamma$ is the corresponding photon spectral index with a value of $1.65$ \citep{2021GCN.30018....1M}. The precession model is calculated with some of the parameters fixed to certain values. The $R$ and $I$ of the SMM are fixed to $12.05 \times 10^{6}\,{\rm cm}$ and $3.33 \times 10^{45}\;{\rm g\;cm^{-2}}$ respectively, which are consistent with the GM1 EoS we have adopted. The suspect collapse time of the SMM in GRB 210514A is $\tau_{\rm c} = t_{\rm obs,b} / (1 + z) = 53\;{\rm s}$ corresponding to the first break in the global lightcurve. Also, according to the observed QPO signature reported in Sec.~\ref{sec:data}, we estimate the period of the precession motion to be $P_{\rm p} = 11 / (1 + z) =7\;{\rm s}$ so that $\Omega_{\rm p} = 2 \pi / P_{\rm p} = 0.90\;{\rm rad/s}$. Subsequently, our model ends up with $7$ free parameters $B_{\rm p}$, $P_{\rm s,0}$, $\alpha_{0}$, $k$, $a$, $b$, and $\Delta t$.
	
The fitting is conducted within the Bayesian statistical inference through the Markov Chain Monte Carlo (MCMC) method. The fitting procedure is implemented with the \texttt{emcee} python module \citep{2013PASP..125..306F} and uniform prior probability distributions are set for each free parameter since we do not have any strong information for them. The parameter boundaries are presented in the Tab.~\ref{tab:boundaries}. The fit results are summarized in Fig.~\ref{fig:MCMC}, which shows that some of the magnetar parameters are well constrained that $B_{\rm p}=6.03^{+0.28}_{-0.28}\times 10^{15}$~G, $\alpha_0=0.34^{+0.02}_{-0.03}$, $k=0.12^{+0.01}_{-0.01}$, and $\Delta t=5.81^{+0.04}_{-0.04}$~s, where the uncertainties are estimated in 1$\sigma$ confidence level. There are three parameters, $P_{\rm s,0}$, $a$, and $b$, that are poorly constrained by our MCMC fitting. For the $P_{\rm s,0}$ parameter, the bad constraint is because the lightcurve data involved in the fitting do not include a distinct break (since we are assuming that the SMM collapses before the spindown time scale $\tau_{\rm sd}$) and the break is supposed to indicate the value of $\tau_{\rm sd}$. As a consequence, the $B_{\rm p}$ and $P_{\rm s,0}$ parameters in our model are in a degenerate state and only one of them can be properly constrained. As for the parameters $a$ and $b$ which determine the $\delta$, they are badly constrained simply because $\delta \approx 1$, meaning that the amplitude of the $\lambda$ barely evolves with times before the SMM collapse. Subsequently, we adopt the median values of these parameters, i.e. $P_{\rm s,0}=1.14$~ms, $a=5.62\times 10^{-5}$, and $b=5.01\times 10^{-3}$, in calculating the best-fit model. The model line and the BAT observations of the first plateau data are presented in Fig.~\ref{fig:lightcurve}. One can observe that our model is roughly consistent with the data, with an estimated spindown time scale $\tau_{\rm sd} = 79.3\;{\rm s}$. The ellipticity of the SMM is estimated to be $\epsilon \approx \Omega_{\rm s,0} / \Omega_{\rm p} = 1.63 \times 10^{-4}$.
}

\section{Discussion}
\label{sec:dis}
{
\subsection{Implications of the delay time $\Delta t$}
\label{subsec:dis_dt}
{
During the lightcurve fitting with our magnetar precession model, we introduce parameter $\Delta t$ to present the delay time between the birth time of the SMM and the BAT trigger time. This parameter is well-constrained with a value of $\Delta t = 5.81\;{\rm s}$, suggesting that the SMM was formed ahead of the prompt emission phase. If our model is correct, there are several plausible explanations for the delay time. To start with,  $\Delta t$ could reflect the energy building-up time of the newborn SMM. Assuming that the SMM is formed in the collapse of a massive star, which is an appropriate guess since GRB 210514A is a long GRB, the magnetic field of the star may be amplified via the dynamo action within it and this process may take few seconds for the magnetic field to reach $B \gtrsim 10^{15}\;{\rm G}$ \citep{2015ApJ...809...39G}. Next, $\Delta t$ could correspond to the time consumption of forming a relativistic jet. The mechanism is that the newborn SMM is very hot, and launches a dirty neutrino-driven wind to prevent a relativistic outflow \citep{2011MNRAS.413.2031M}. The jet could also been formed through accretion with a similar mechanism to that of a BH, which is tapping the energy accumulated in a hyper-accreting disk. Only, the cooling of the disk is predicted to be more efficient than that of the BH \citep{2008ApJ...683..329Z} and the jet existence time is limited by the accreted masses, in case that the SMM would not collapse into a BH immediately. Plus, the stable jet needs to break out of the stellar envelope left by the progenitor once it is produced, so that its emission can reach earth observers. During this process, the jet would be re-collimated to smaller open angles if the central engine is keeping active, and such a process may last for several to tens of seconds depending on the accretion rate and the equation of state of the magnetar.
}

\subsection{Alternative models}
\label{subsec:dis_mdl}
{
In this section, we discuss alternative models capable of producing the curious double-plateau and steep decay lightcurve structures of GRB 210514A. Regarding the steep decay segments, they are frequently observed not only in the X-ray band but also in the gamma-ray band of GRBs. The most commonly seen explanation for them is that they are the ``tail" of the GRB prompt emission phase \citep{2005ApJ...635L.133B}. When the central engine stops abruptly or turns off with a steeper temporal slope than the observed decline slope, a steep decay can be caused via the curvature effect. This effect is that the progressively fainter prompt photons from increasingly higher latitudes concerning the observer's line-of-sight arrive with time delays since they travel longer distances than photons from low latitudes. Theoretical calculations suggest that lightcurve decay slope $\hat{\alpha}$ due to the curvature effect is related to the spectral index $\hat{\beta}$ as $\hat{\alpha} = 2 + \hat{\beta}$  \citep{2000ApJ...541L..51K}, and $\hat{\beta} = 1$ is a typical value that is often used in estimating $\hat{\alpha}$ assuming that the prompt emission is produced in a shell moving with a relativistic speed. Note that the study also suggests that the decay slope could be much steeper than the prediction of  $\hat{\alpha} = 2 + \hat{\beta}$ if the shell is accelerating \citep{2015ApJ...808...33U}. The X-ray plateau, on the other hand, is usually followed by a normal decay segment with a characteristic slope $\sim -1.3$. It is conventionally interpreted in the framework of the external shock models, where the central engine provides a continuous energy injection into the shocks \citep{2018MNRAS.480.4402L}. In this scenario, the injected energy is ultimately produced via the accretion process if the central engine is a BH (e.g. \citealt{2018MNRAS.475..266Z}).

In the case of GRB 210514A, if a BH central engine is present, the {\it Swift}-XRT lightcurve observations could be explained as a result of the sudden cease and the late-time activity of the central engine. Intriguingly, \cite{2015ApJ...806..205D} propose a collapsar model in that no energy injection is needed for generating such a plateau, which can be formed naturally in the jet propagation. In their model, the outflow from a BH central engine forms a top-heavy jet, i.e. a highly relativistic jet core is preceded by a heavy, baryon-loaded outer shell. This outer shell is massive enough to significantly decelerate the jet core upon collision, resulting in a less relativistic amalgamated jet. The external shocks of the jet naturally produce a plateau with a slope $\sim 0.25$. During the decelerated phase of the jet core, the collision procedure could generate a steep decay of slope $\sim -8$ before the plateau. This model can also provide a reasonable explanation for GRB 210514A. However, in the models discussed above, the QPO signal observed by the BAT is rather awkward. There are a few mechanisms concerning a BH central engine that may be able to produce the signal, taking the jet precession motion as an example, but the physics would be much more complicated compared to that of the SMM model. Or, the observed QPO signal could merely be a coincidence of internal shocks since the pulses shown in the prompt emission phase originate from the collisions of shells emitted successively by the central engine.
}

\section{Conclusions}
\label{sec:con}
{
In this paper, we performed a detailed analysis of GRB 210514A. This burst is exceptionally rare not only because it has an internal plateau, but also because it is the only burst in our searches that has a QPO signal in the prompt emission phase observed by {\it Swift}-BAT. The QPO signal has a period $\sim 11$~s and a confidence level over 3$\sigma$. Inspired by this, we conduct a comprehensive analysis of the gamma-ray and X-ray features observed in GRB 210514A and explore their possible origins. The lightcurve of GRB 210514A is unusual, featuring two distinct plateaus that deviate from typical GRB lightcurves. We propose that the first plateau arises from the magnetic DR emitted by a newborn SMM. The subsequent sharp decay in the lightcurve suggests the collapse of the SMM into a BH. We consider the possibility of fall-back accretion onto the newborn BH, potentially explaining the second plateau via the BZ process \citep{1977MNRAS.179..433B,2018MNRAS.475..266Z}. This model aligns well with the observed data. The QPO signal can be well presented with the DR model when taking the magnetar precession motion into account. The derived precession angular velocity is $\Omega_{\rm p} = 0.90$~rad/s. Adopting the EOS GM1 and assuming $\eta_{X}/f_{\rm b}=1$, we derive the parameters, including the magnetic field strength $B_{\rm p}= 6.03\times10^{15}$~G, the initial spin period $P_{\rm s,0}= 1.14$~ms, the initial inclination angle $\alpha_{0} = 0.34$, $k = 0.12$, $a = 5.62\times 10^{-5}$, $b = 5.01\times10^{-3}$, and the delay time $\Delta t=5.81$~s. The derived best-fit parameters and the results of our MCMC analysis suggest that the magnetar precession motion is steady with $\delta \approx 1$, the estimated spindown time scale is $\tau_{\rm sd} = 79.3\;{\rm s}$, which is consistent with our model assumption that the SMM collapses before its spins down. Hopefully, future observations will find more similar GRBs, and by studying them, our knowledge about the central engine of those fascinating events will be improved.

One caveat should be addressed that the pseudo-redshift of GRB 210514A adopted in this study, derived from the $L-T-E$ correlation \citep{2023ApJ...943..126D}, may introduce a degree of uncertainty. The quest to determine the pseudo-redshift of GRBs has seen researchers employing various empirical correlations \citep{2002A&A...390...81A, 2003A&A...407L...1A, 2004ApJ...609..935Y, 2005ApJ...633..611L, 2011ApJ...730..135D, 2013ApJ...772L...8T, 2013MNRAS.431.1398T}. For example, \cite{2003A&A...407L...1A} utilized the Amati relation to estimate pseudo-redshifts for 18 GRBs detected by HETE/FREGATE. Their estimate for the redshift of GRB 980326 at 1.05 corroborated the observational estimate of $z=1$, inferred from a supernova bump in the late afterglow lightcurve \citep{1999Natur.401..453B}. Similarly, the pseudo-redshift for GRB 000214 was calculated at 0.39, aligning with the estimated $z=0.42$ from the observation of an iron line in its X-ray afterglow \citep{2000ApJ...545L..39A}. \cite{2013MNRAS.431.1398T} determined pseudo-redshifts for 71 bright sGRBs by employing a robust correlation between the peak isotropic luminosity in the observed frame and the spectral peak energy in the rest-frame. Their results notably revealed that the pseudo-redshifts of sGRBs tend to cluster around $z=1.05$, which is supported by the merger scenario of sGRBs. \cite{2019arXiv190900887S} also employed a relation between the non-thermal component's peak energy and the luminosity to constrain the pseudo-redshifts of GRBs \citep{2015ApJ...807..148G}. In doing so, they were able to infer pseudo-redshifts for GRBs 080916C, 090926A, and 150314A that closely aligned with the reported redshifts. These studies collectively suggest that the pseudo-redshifts serve as reasonable estimates for the distances of GRBs, especially in cases where spectral redshifts are not available.
}
\section{Acknowledgements}

We appreciate the very helpful comments and suggestions from the referee. We thank En-Wei Liang for the useful discussion. We acknowledge the use of the public data from the {\em Swift} data archive and the UK {\em Swift} Science Data Center.

{}

\clearpage

\begin{table}
\center
\caption{The best fitting parameters of the lightcurve}
\begin{tabular}{cccccc}
\hline
\hline
 &  $p_{\rm 1}$ & $p_{\rm 2}$ & $t_{\rm obs,b}$ & reduced-$\chi^{2}$ \\
 &      &     & s               &                    \\
\hline
$t_{\rm obs} \sim [0, 100]\;{\rm s}$  & $0.21 \pm 0.05$ & $6.55 \pm 0.80$ & $83 \pm 2$     & 0.47 \\
$t_{\rm obs} > 100\;{\rm s}$           & $0.01 \pm 0.04$ & $1.30 \pm 0.03$ & $1302 \pm 100$ & 0.95 \\
\hline
\hline
\end{tabular}
\label{tab:fitting}
\end{table}

\begin{table}
\center
\caption{Boundaries of Parameters}
\begin{tabular}{cccccc}
\hline
\hline
Parameters & Unit & Lower Limit & Upper Limit \\
\hline
${\rm lg}(B_{\rm p})$ & ${\rm G}$ & $15.0$ & $16.0$ \\
$P_{\rm s,0}$ & ${\rm ms}$ & $0.99$ & $1.2$ \\
$\alpha_{\rm 0}$ & ${\rm rad}$ & $0.01$ & $1.0$ \\
$k$ & $\rm rad$ & $0.1$ & $1.0$ \\
${\rm lg}(a)$ & -- & $-5.0$ & $0.0$ \\
${\rm lg}(b)$ & -- & $-5.0$ & $0.0$ \\
$\Delta t$ & ${\rm s}$ & $0.0$ & $10.0$ \\
\hline
\hline
\end{tabular}
\label{tab:boundaries}
\end{table}

\begin{figure}
\center
\includegraphics[angle=0,scale=0.30]{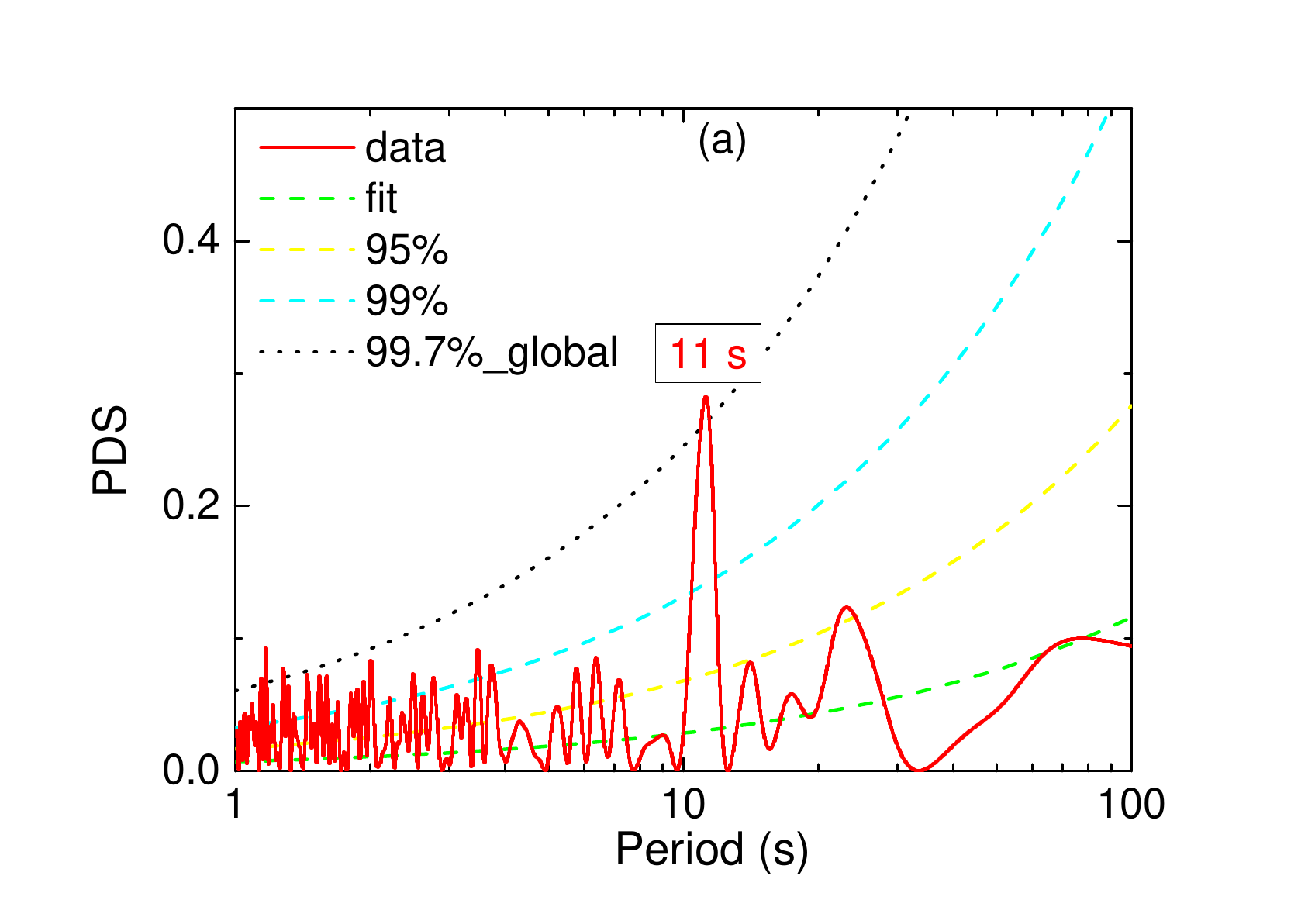}
\includegraphics[angle=0,scale=0.30]{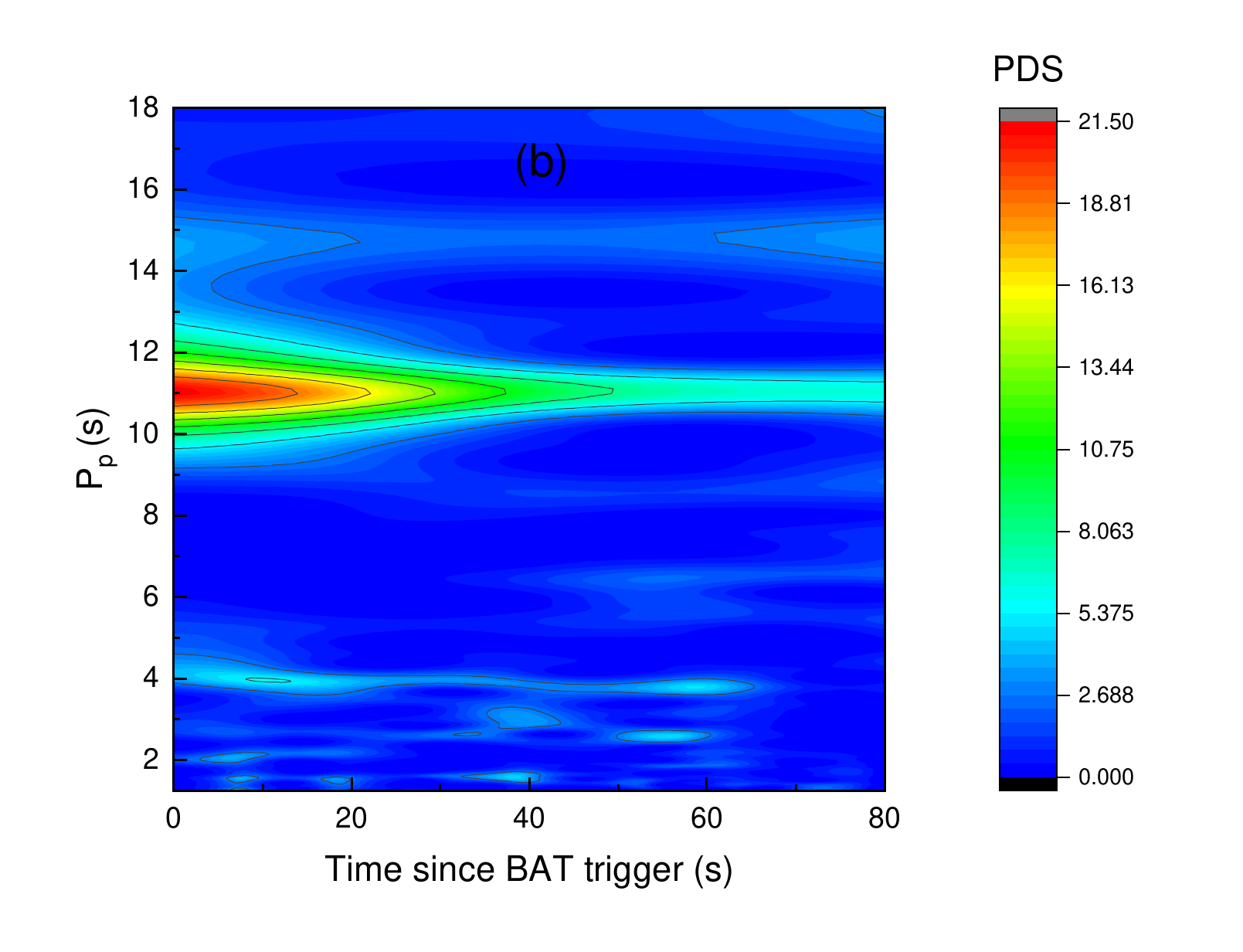}
\center\caption{{\em Left panel}--- PDS calculated with the LSP algorithm for the observed (red line) lightcurves. The best-fit red noise spectrum is shown as a green dashed line. Single-frequency 95\% and 99\% confidence level lines are reported by yellow and Cyan dashed lines, and the global 99.7\% false-alarm levels of the PL model are shown as black dotted lines.  {\em Right panel}--- Power-density spectrum (PDS) derived from our time-frequency domain analysis with the WWZ algorithm for the gamma/X-ray lightcurve of GRB 210514A. }
\label{fig:QPO}
\end{figure}

\begin{figure}
\center
\includegraphics[angle=0,scale=0.35]{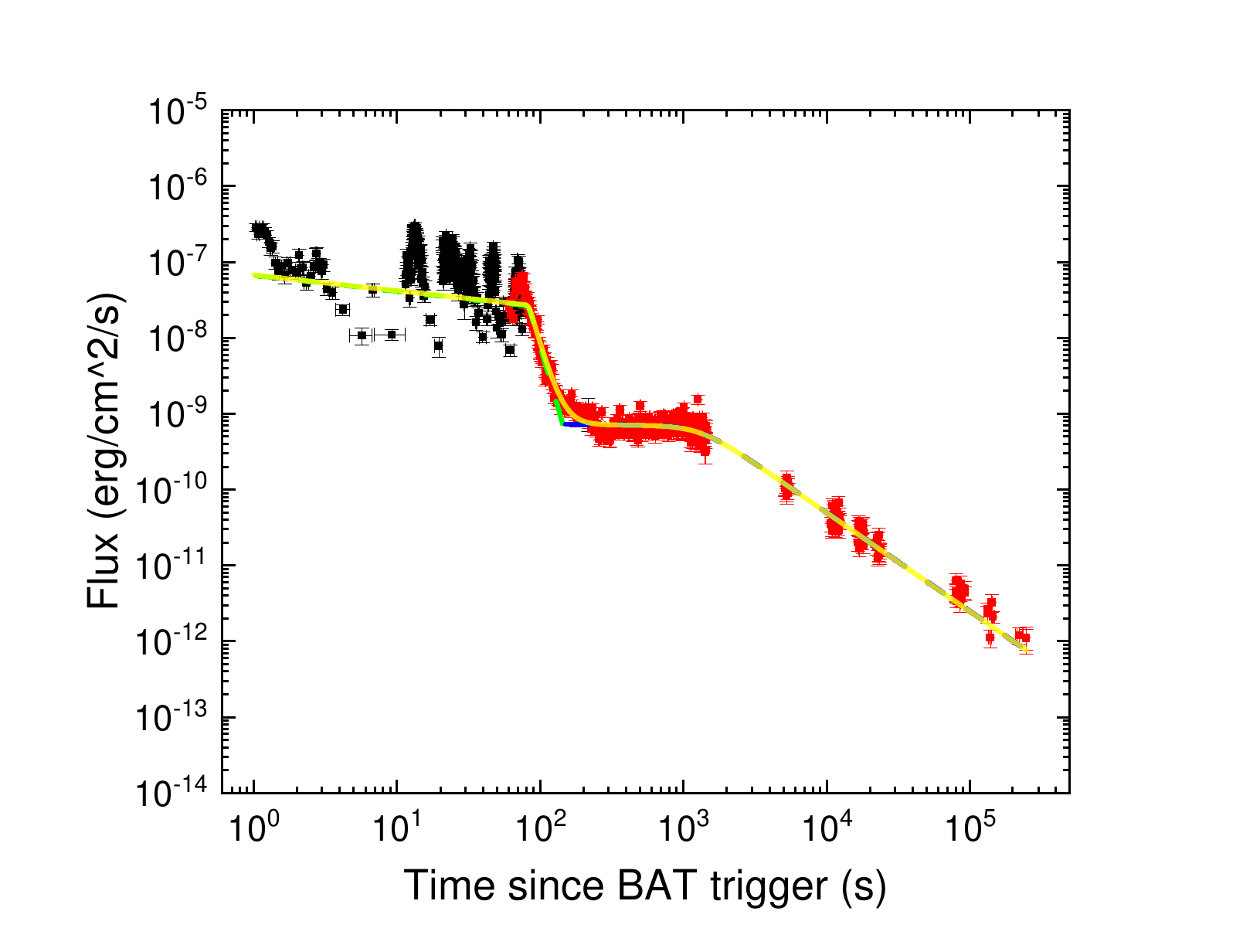}
\center\caption{Joint BAT (black dots) and XRT afterglow (red dots) lightcurves of GRB 210514A. The global feature of the gamma/X-ray lightcurve is depicted with two smooth broken power-law functions as marked with a yellow line. The parameters for these two parts, denoted as ($p_{1}$, $p_{2}$, $t_{\rm obs,b}$), are (0.21$\pm$0.05 , 6.55$\pm$0.80, 83$\pm$2~s) for the first part and (0.01$\pm$0.04, 1.30$\pm$0.03, 1302$\pm$100~s) for the second part, respectively.}
\label{fig:lightcurve}
\end{figure}

\begin{figure}
\center
\includegraphics[angle=0,scale=0.35]{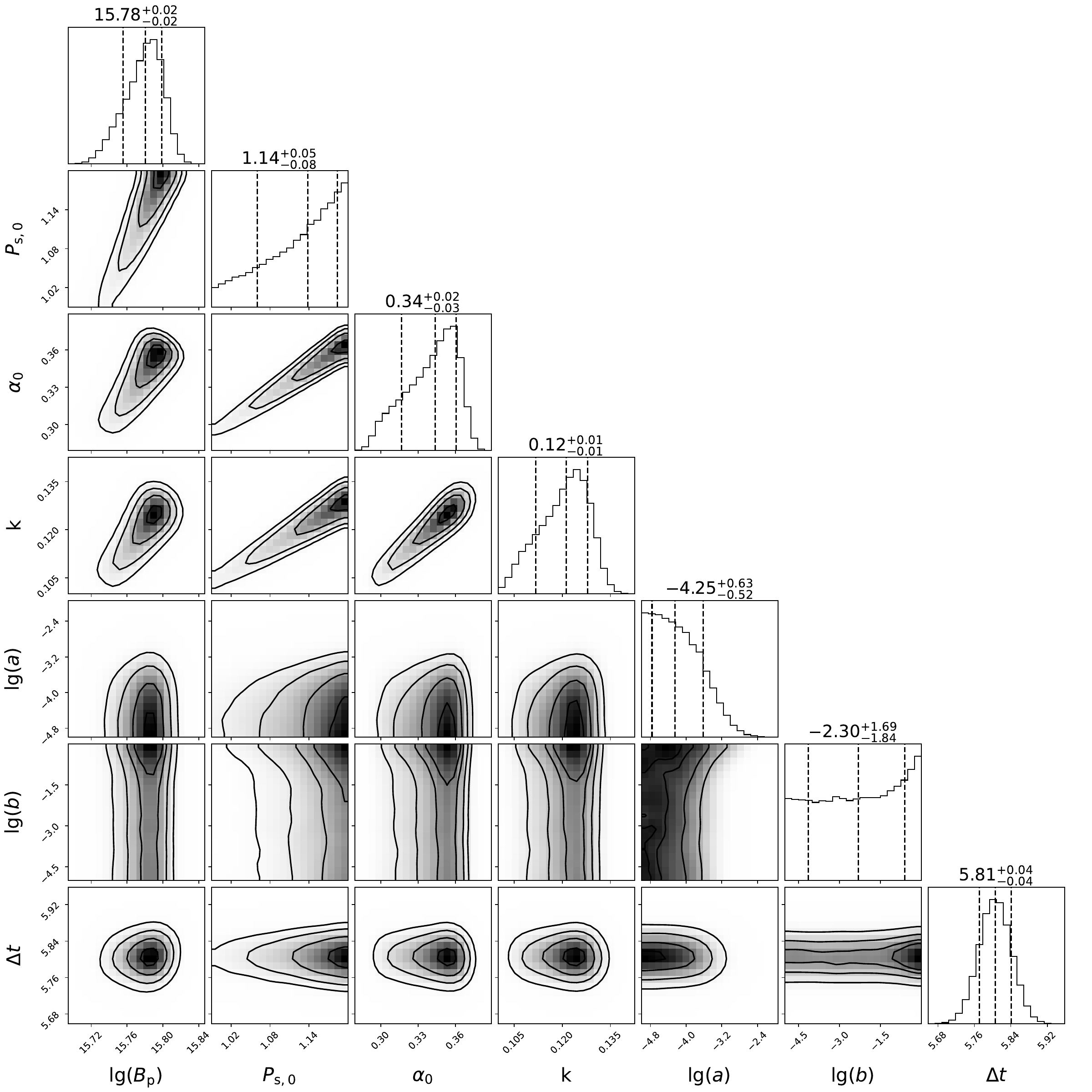}
\center\caption{Probability contours of the model parameters derived from our MCMC fit for GRB 210514A. The vertical dashed lines mark the 1$\sigma$ confidence level regions centering at their median probabilities.}
\label{fig:MCMC}
\end{figure}

\begin{figure}
\center
\includegraphics[angle=0,scale=0.35]{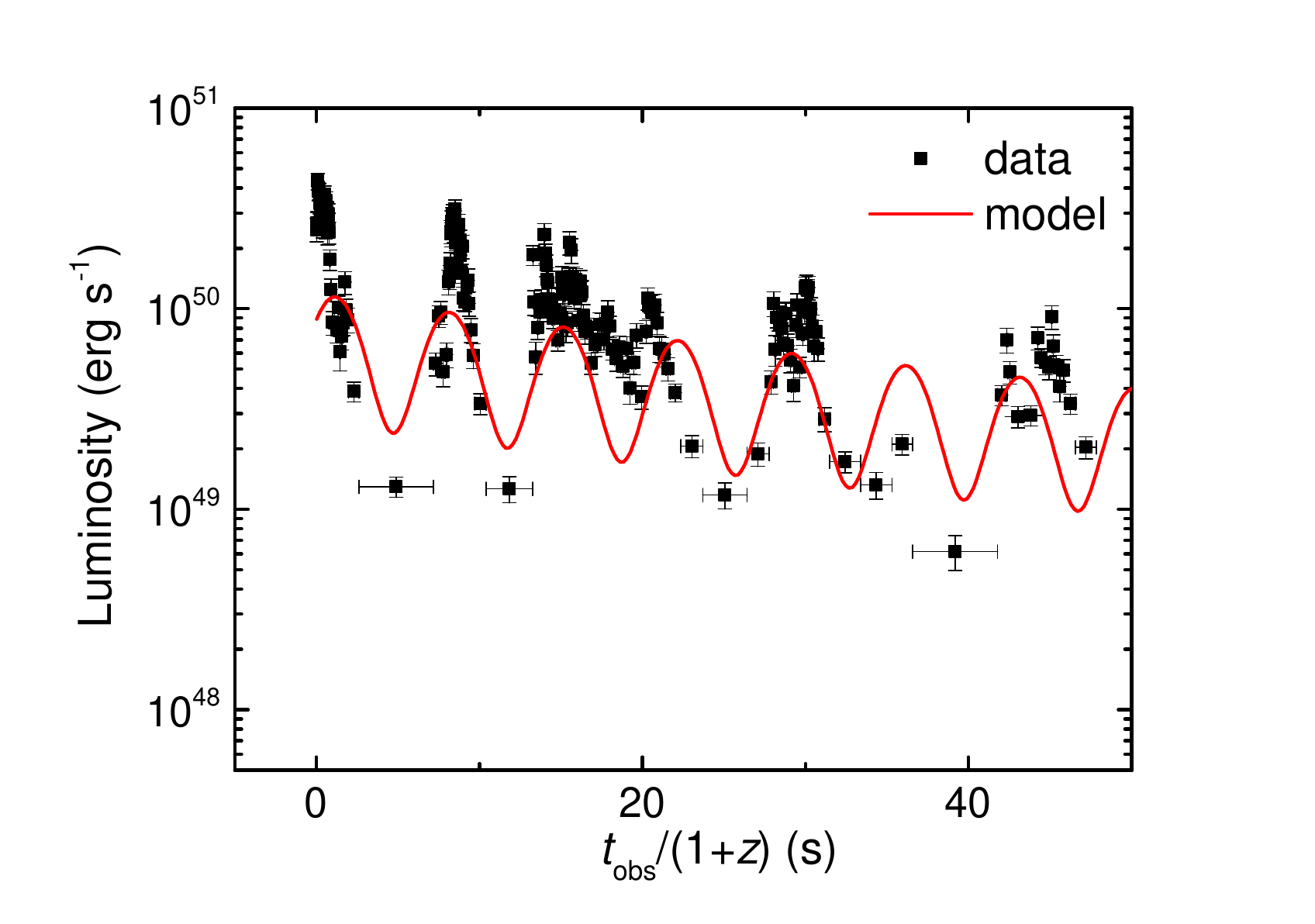}
\center\caption{The model lightcurve (red line) derived from our best model fit in comparison with the observed lightcurve of GRB 210514A (black dots).}
\label{fig:precession}
\end{figure}

\end{document}